\documentclass[amsmath,amssymb,nofootinbib,superscriptaddress]{revtex4-1}
\pdfoutput=1

\usepackage{aas_macros,esint,etoolbox,graphicx,wallpaper}
\usepackage[colorlinks=true]{hyperref}

\let\originalleft\left
\let\originalright\right
\renewcommand{\left}{\mathopen{}\mathclose\bgroup\originalleft}
\renewcommand{\right}{\aftergroup\egroup\originalright}
\mathcode`\*="8000
{\catcode`\*=\active\gdef*{\mathclose{}\,\mathopen{}}}

\newcommand{\ab}[1]{\left|#1\right|}

\newcommand{\br}[1]{\left[#1\right]}
\newcommand{\cu}[1]{\left\{#1\right\}}
\newcommand{\pa}[1]{\left(#1\right)}

\newcommand{\ed}{\mathop{}\!\mathrm{d}}
\newcommand{\pd}{\mathop{}\!\partial}
\renewcommand{\L}{\mathcal{L}}
\renewcommand{\O}[1]{\mathcal{O}\pa{#1}}
\DeclareMathOperator\dn{dn}
\DeclareMathOperator\sn{sn}
\DeclareMathOperator\sign{sign}

\begin{document}

\title{\Huge{Polarization Whorls from M87* at the Event Horizon Telescope}
\vspace{30pt}}

\email{lupsasca@fas.harvard.edu\\
kapec@ias.edu\\
yshi@g.harvard.edu\\
dgates@g.harvard.edu\\
strominger@physics.harvard.edu}

\author{Alexandru Lupsasca}
\affiliation{Center for the Fundamental Laws of Nature, Harvard University, Cambridge, MA 02138, USA}
\affiliation{Society of Fellows, Harvard University, Cambridge, MA 02138, USA}
\author{Daniel Kapec}
\affiliation{Center for the Fundamental Laws of Nature, Harvard University, Cambridge, MA 02138, USA}
\affiliation{School of Natural Sciences, Institute for Advanced Study, Princeton, NJ 08540, USA}
\author{Yichen Shi}
\affiliation{Center for the Fundamental Laws of Nature, Harvard University, Cambridge, MA 02138, USA}
\author{Delilah E.~A. Gates}
\affiliation{Center for the Fundamental Laws of Nature, Harvard University, Cambridge, MA 02138, USA}
\author{Andrew Strominger}
\affiliation{Center for the Fundamental Laws of Nature, Harvard University, Cambridge, MA 02138, USA}

\begin{abstract}
\vspace{60pt}
\large{The Event Horizon Telescope (EHT) is expected to soon produce polarimetric images of the supermassive black hole at the center of the neighboring galaxy M87. There are indications that this black hole is rapidly spinning. General relativity predicts that such a high-spin black hole has an emergent conformal symmetry near its event horizon. In this paper, we use this symmetry to analytically predict the polarized near-horizon emissions to be seen at the EHT and find a distinctive pattern of whorls aligned with the spin.}
\end{abstract}

\maketitle

\color{red}
\tableofcontents
\color{black}
\ThisCenterWallPaper{1}{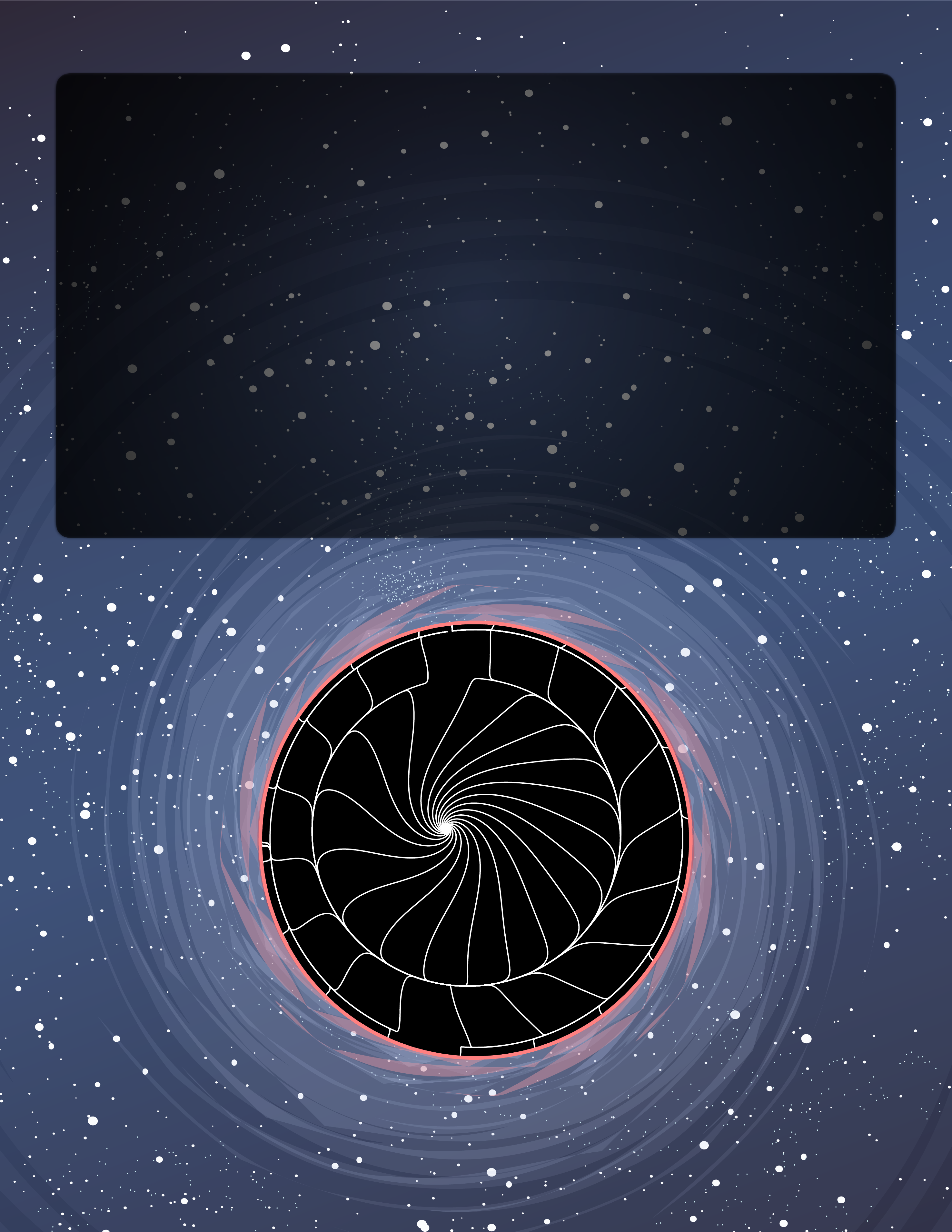}

\clearpage

\hfill

\section{Introduction}

Observational black hole astronomy is entering an exciting era of rapid progress. With its celebrated detection of gravitational waves from a binary black hole merger, the LIGO collaboration has provided scientists with a radically new tool to study black holes \cite{Abbott2016a,Abbott2016b,Abbott2017a,Abbott2017b,Abbott2017c}. Meanwhile, in the arena of electromagnetic astronomy, the Event Horizon Telescope (EHT) has recently delivered the first-ever up-close pictures of the supermassive black hole at the center of the galaxy Messier 87 (M87*) \cite{EHT2019a,EHT2019b,EHT2019c,EHT2019d,EHT2019d,EHT2019e} and will soon release images of the black hole at the center of our own galaxy (Sagittarius A*) \cite{Doeleman2017}. As its capabilities improve, the EHT will eventually resolve the near-horizon regions of these black holes with a few dozen pixels at the horizon scale. This offers an unprecedented opportunity for theorists to make predictions: \textit{What will the images look like?}

The data collected by the EHT will provide a wealth of information about the electromagnetic emissions from the black hole's vicinity. The optical appearance of the black hole is largely determined by the brightness of the surrounding emission region, and has been the focus of intense investigation \cite{Broderick2006,Doeleman2008a,Doeleman2008b,Doeleman2009,Doeleman2012,Johnson2015}. The EHT will also measure the polarization of incident light, which is expected to carry important information about dynamics in the region surrounding the black hole \cite{Bromley2001,Shcherbakov2012,Shcherbakov2013,Chael2016,Gold2017,Akiyama2017,Moscibrodzka2017,Birdi2018,Pihajoki2018,JimenezRosales2018}. The present work concerns this \textit{polarimetric image}, which has received comparatively less attention.

The determination of the optical and polarimetric image of a generic black hole is in general an arduous task, as one must account for a multitude of complex astrophysical effects. This task is further complicated by the need to make numerous assumptions regarding the black hole's environment, such as the surrounding matter distribution and its radiative properties. As a result, analytic computation is in most cases completely infeasible. Instead, extensive numerical simulation is required in order to properly account for myriad physical ingredients. The proliferation of tunable parameters and model-dependent assumptions can make it difficult to extract physical predictions or intuition from numerics.

However, in the special case of a high-spin black hole, a dramatic simplification takes place. Certain features exhibit universal `critical' behavior that is independent of the detailed assumptions and parameters entering the model. This occurs because general relativity dictates that a rapidly spinning black hole develops an emergent conformal symmetry in its near-horizon region \cite{Bardeen1973,Bardeen1999,Guica2009}. This conformal symmetry imposes strong constraints on fields and matter near the horizon with observable astrophysical consequences. Over the past several years, this symmetry has been judiciously exploited in order to analytically compute several otherwise intractable\footnote{Numerical analyses are particularly problematic in this regime due to the critical behavior, since the (near-)extreme Kerr metric (in any coordinate system) fails to adequately resolve near-horizon physics (see Sec.~III of Ref.~\cite{Kapec2020} for an astrophysically oriented review of these issues). Hence, the analytic and numeric approaches have complementary domains of applicability.} astrophysically relevant processes \cite{Andersson2000,Glampedakis2001,Yang2013,Porfyriadis2014a,Porfyriadis2014b,Hadar2015,Gralla2016a,Gralla2016b,Burko2016,Hadar2017,Compere2017,Kapec2020,Compere2020}.

Fortunately, there are some indications that M87* could be just such a high-spin black hole. Black holes with pronounced jets have often been found to have high spins, and the spin of M87* was argued to be within 2\% of criticality in Ref.~\cite{Feng2017}. In this paper, we assume that M87* is indeed in the high-spin regime, and exploit its emergent conformal symmetry to derive a universal prediction for the polarization profile of its near-horizon emissions. The simplifying nature of the extremal limit allows us to perform our calculations analytically, and we obtain striking polarimetric images of the near-horizon emissions. Their signal lies entirely within the shadow and thus is not obscured by emissions from behind the black hole. The polarization lines form a distinctive `whorl' that spirals into a central point inside the shadow. An infinite series of subsidiary whorls also appear nearer the edge of the shadow, arising from photons that librate around the black hole multiple times before escaping to infinity. 

Local emissions near the horizon of a black hole have large gravitational redshift relative to a distant observer such as the EHT. Our prediction for the polarimetric image of M87* assumes that there is some physical process in its near-horizon region capable of producing photons whose redshifted energy would lie in the observation band of the EHT, and that is moreover invariant under the same emergent symmetries as the background geometry. The latter assumption is standard for the more familiar cases of axisymmetry and time-translation symmetry. Even if these assumptions hold, and the EHT attains the hoped-for resolution, the signal might not be observable if it is obscured by emissions anywhere between the near-horizon region and the telescope. Finally, Faraday rotation could potentially distort the predicted image, though such propagation effects are attenuated for highly energetic photons.

The outline of this paper is as follows. In Section~\ref{sec:Summary}, we summarize our findings and present our prediction for the polarimetric image of M87* (Fig.~\ref{fig:M87}). Section~\ref{sec:GeometricOptics} collects important results about the Kerr geometry and reviews the geometric optics approximation. Section~\ref{sec:Computation} outlines our calculation of the polarimetric image. Various technical steps are relegated to the appendices.

\section{Summary}
\label{sec:Summary}

Our setup consists of an observer located at a large distance $r_o\to\infty$ from a rotating (Kerr) black hole of mass $M$ and spin $J$, stationed at a fixed polar angle $\theta_o$ relative to the black hole's axis of symmetry. For the sake of simplicity, we perform our calculations for a maximally spinning ($\ab{J}=M^2$) black hole, but we expect our results to remain applicable for very rapidly spinning black holes such as M87*, since they share the same near-horizon geometry.

In the geometric optics approximation, which is valid for the highly energetic emissions considered here, photons propagate along null geodesics from the near-horizon region towards the observer. Our computation of polarimetric images therefore involves two separate steps, both of which are rendered analytically tractable by simplifications that arise in the high-spin regime. First, we must determine the polarization properties of the radiation in the near-horizon region of the black hole. The assumption that this polarization profile respects the emergent conformal symmetry of the throat region singles out a unique polarization distribution. In order to translate this near-horizon data into a polarimetric image for the far observer, we must then parallel transport each photon's polarization vector out to the observer's screen in the far region. This involves solving the geodesic equation for the photon, as well as the parallel propagation of the polarization vector along the geodesic.

The geodesic motion of a particle in the Kerr spacetime is characterized by its invariant mass $\mu$, energy $\omega$, angular momentum about the axis of symmetry $\ell$, and Carter constant $k$ [Eq.~\eqref{eq:KerrConserved}]. For null geodesics, the energy scales out but is replaced by a new conserved quantity: the Penrose-Walker constant $\kappa$, which can be used to parallel transport the polarization vector of the light ray [Eq.~\eqref{eq:KerrPW}]. The photons of interest here are emitted from the vicinity of the black hole and impinge upon the screen of the observer, which we label with Cartesian coordinates $(\alpha_o,\beta_o)$. These incident parameters are determined by the conserved quantities as follows [Eq.~\eqref{eq:ScreenCoordinates}]:
\begin{align}
	\alpha_o=-\frac{\ell}{\omega\sin{\theta_o}},\qquad
	\beta_o=\pm_o\frac{\sqrt{k-\pa{\ell\csc{\theta_o}-a\omega\sin{\theta_o}}^2}}{\omega},
\end{align}
with the sign $\pm_o$ equal to that of the photon's momentum in the $\theta$-direction when it reaches the observer.

Since the source polarization profile is stationary and axisymmetric by assumption, we can ignore the $(t,\phi)$ components of the geodesic equation and only need to consider its poloidal $(r,\theta)$ part. In terms of a radial coordinate $R$ that vanishes on the horizon, a photon emitted from a source located at $(R_s\sim0,\theta_s)$ reaches the distant observer at $(R_o\to\infty,\theta_o)$ provided that $I_r=G_\theta$, where
\begin{align}
	I_r=\fint_{R_s}^{R_o}\frac{\ed R}{\pm\sqrt{R^4+4R^3+\pa{7-q^2-\lambda^2}R^2+4(2-\lambda) R+(2-\lambda)^2}},\qquad
	G_\theta=\fint_{\theta_s}^{\theta_o}\frac{\ed\theta}{\pm\sqrt{q^2+\cos^2{\theta}-\lambda^2\cot^2{\theta}}}.
\end{align}
Here, $(\lambda,q)$ are the dimensionless, energy-rescaled angular momentum and Carter constant, respectively [Eq.~\eqref{eq:DimensionlessConstants}]. The slash on the integrals indicates that they are to be evaluated along the geodesic, with the signs of the integrands chosen to match those of $\ed r$ and $\ed\theta$, respectively. Note that in Boyer-Lindquist coordinates (rescaled or not), a null geodesic can encounter at most one turning point in its radial motion. On the other hand, it can undergo multiple librations (oscillations in the polar motion) along its trajectory from the throat region to the observer.

It is well-known that a black hole casts a shadow on the observer's screen whose edge corresponds to light rays that asymptote to bound photon orbits [Eqs.~\eqref{eq:Shadow}--\eqref{eq:ExtremalShadow}]. We find that near-horizon emissions are confined to appear within this shadow---that is, they cannot be obscured by bright sources located far behind the black hole [App.~\ref{app:GeodesicEquation}]. Moreover, each point $(\alpha_o,\beta_o)$ inside the shadow on the observer's screen corresponds to a geodesic that traverses the near-horizon region at a fixed angle $\theta_s$. Solving $I_r=G_\theta$ for this geodesic allows us to determine its angle $\theta_s$ of emission from the throat. This procedure is described in detail in App.~\ref{app:GeodesicEquation}, where we derive a remarkably simple analytic formula for $\theta_s$ as a function of the observer parameters [Eq.~\eqref{eq:SourceAngle}]:
\begin{align}
	\theta_s=\arccos\br{\pm_o\sqrt{u_+}\sn\pa{\frac{\sqrt{-u_-}}{\sign(-u_-)}I_r\pm_oF\pa{\arcsin\pa{\frac{\cos{\theta_o}}{\sqrt{u_+}}}\bigg|\frac{u_+}{u_-}}\bigg|\frac{u_+}{u_-}}}.
\end{align}
Here, $F$ denotes the incomplete elliptic integral of the first kind and
\begin{align}
	u_{\pm}=\frac{1}{2}\pa{1-q^2-\lambda^2}\pm\sqrt{\frac{1}{4}\pa{1-q^2-\lambda^2}^2+q^2}.
\end{align}

Photons impinging on the observer's screen register a polarization defined by the direction of oscillation of the transverse electric field. For our symmetric choice of source polarization, this direction is given by [Eqs.~\eqref{eq:PolarizationAngle}$\,\&\,$\eqref{eq:ObservedPolarization}]
\begin{align}
	\label{eq:Prediction}
	\vec{\mathcal{E}}=\frac{1}{\pa{\beta_o^2+\gamma_o^2}}\pa{\beta_o\gamma_s-\gamma_o\beta_s,\beta_o\beta_s+\gamma_o\gamma_s},
\end{align}
where the prefactor is fixed by the normalization condition $\vec{\mathcal{E}}\cdot\vec{\mathcal{E}}=1$ (though its overall sign is irrelevant), and\footnote{In the limit $M\to0$ in which the black hole disappears, $\gamma_i\to-\alpha_i$ and $\alpha_s\to\alpha_o$ since geodesics in flat spacetime are straight lines. In that case, $\vec{\mathcal{E}}\to(0,1)$ and our symmetric polarization pattern reduces to a uniform electric field.}
\begin{align}
	\gamma_i=-(\alpha_i+M\sin{\theta_i}),\qquad
	\alpha_s=\frac{\sin{\theta_o}}{\sin{\theta_s}}\alpha_o,\qquad
	\beta_s=(-1)^m\beta_o\sqrt{1+\frac{\gamma_o^2-\gamma_s^2}{\beta_o^2}}.
\end{align}
Here, $m$ denotes the number of librations of the geodesic connecting the source located at $(R_s\sim0,\theta_s)$ to the observer. It can be eliminated from Eq.~\eqref{eq:Prediction} using the procedure outlined in App.~\ref{app:GeodesicEquation}.

\begin{figure}
	\centering
	\includegraphics[width=.49\textwidth]{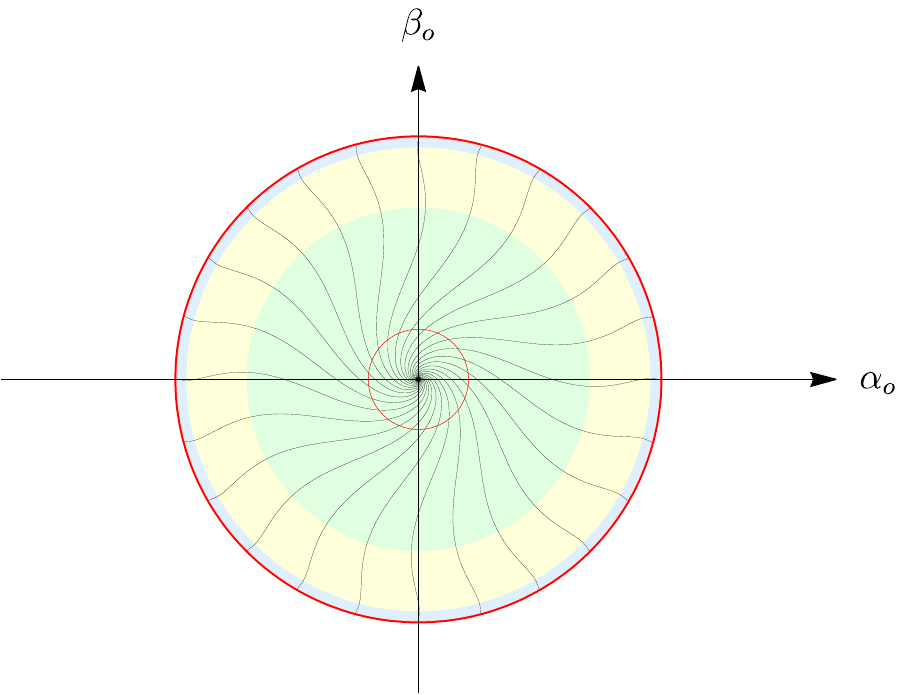}\quad
	\includegraphics[width=.49\textwidth]{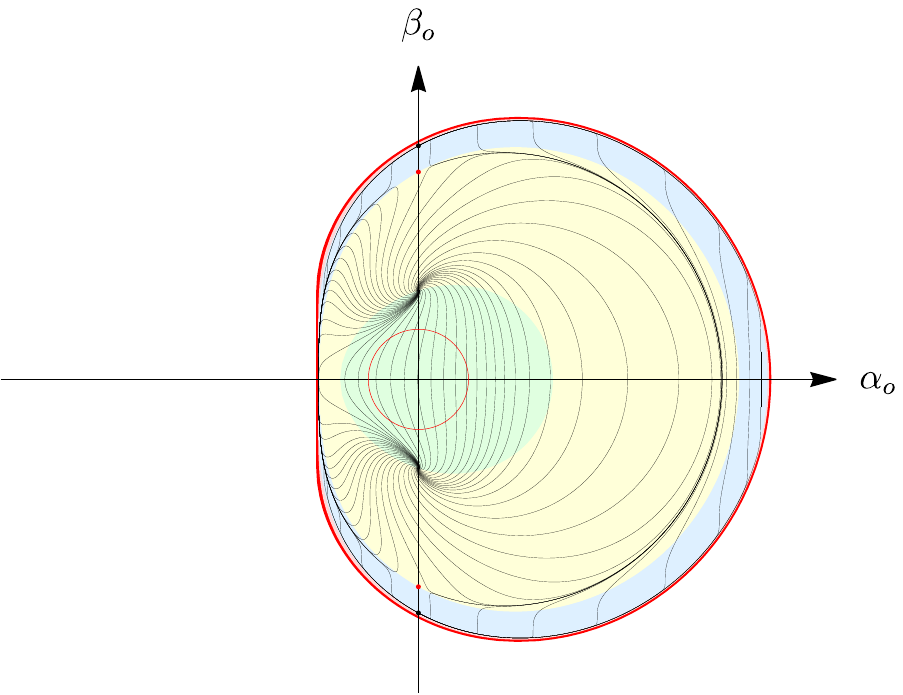}
	\caption{Polarimetric image observed by a `face-on' observer at the pole ($\theta_o=0^\circ$) and by an `edge-on' observer in the equatorial plane ($\theta_o=90^\circ$). The black lines are integral curves of the electric field \eqref{eq:Prediction}. The inner red circle is the locus $\alpha_o^2+\beta_o^2=M^2$, corresponding to the would-be location of the event horizon, while the outer red contour is the edge of the shadow. The black hole's spin is counterclockwise about the origin in the left image, and right-handed about the $\beta_o$-axis in the right image. The left image, which is circularly symmetric in accordance with the axisymmetry of the Kerr black hole, depicts a swirling pattern around the pole that resembles a hair whorl. The right image, which is reflection symmetric in accordance with the equatorial symmetry of the Kerr black hole, contains two whorls because gravity bends light from both the northern and southern poles back towards the observer. The colors indicate the number $m$ of angular turning points of the geodesic connecting the source to the observer: $m=0$ (direct) light appears in the green region, while the yellow, blue, and red regions correspond to relativistic images with $m=1,2$, and $3$, respectively. There are infinitely many more regions (which correspond to geodesics with diverging $m$) bunched up near the edge of the shadow (which corresponds to asymptotically bound photon orbits) but their area rapidly decreases with $m$ and they become indiscernible past $m>3$.}
	\label{fig:FaceOnEdgeOn}
\end{figure}

Example plots of the polarimetric image \eqref{eq:Prediction} at $\theta_o=0^\circ$ and $\theta_o=90^\circ$ are depicted in Fig.~\ref{fig:FaceOnEdgeOn}. Both images display striking features. In the face-on ($\theta_o=0^\circ$) case, the integral curves of the electric field, which describe the direction of local linear polarization, assume a `whorling' pattern. They spiral outwards from the origin, which is an optical image of the north pole of the black hole, and pass through successive images of the poles (appearing as an infinite series of concentric circles converging to the shadow edge), with the connecting geodesic increasing its number of angular turning points by one after each pass.

The image seen by an edge-on observer ($\theta_o=90^\circ$) is considerably less symmetric, and presents qualitatively new features. The integral curves of the local polarization `whorl' around special points on the $\alpha_o=0$ axis corresponding to strongly lensed images of the north and south poles. Successive images of the poles alternate between accumulation points and repulsion points (depicted as black and red dots, respectively). Domain walls corresponding to local extrema of the map $\theta_s(\alpha_o,\beta_o)$ control the pattern farther away from the origin, and accumulate near the shadow of the black hole. This pattern repeats indefinitely, with an infinite set of whorls demarcating an infinite number of coverings of the horizon of the black hole!

\begin{figure}
	\includegraphics[width=.8\textwidth]{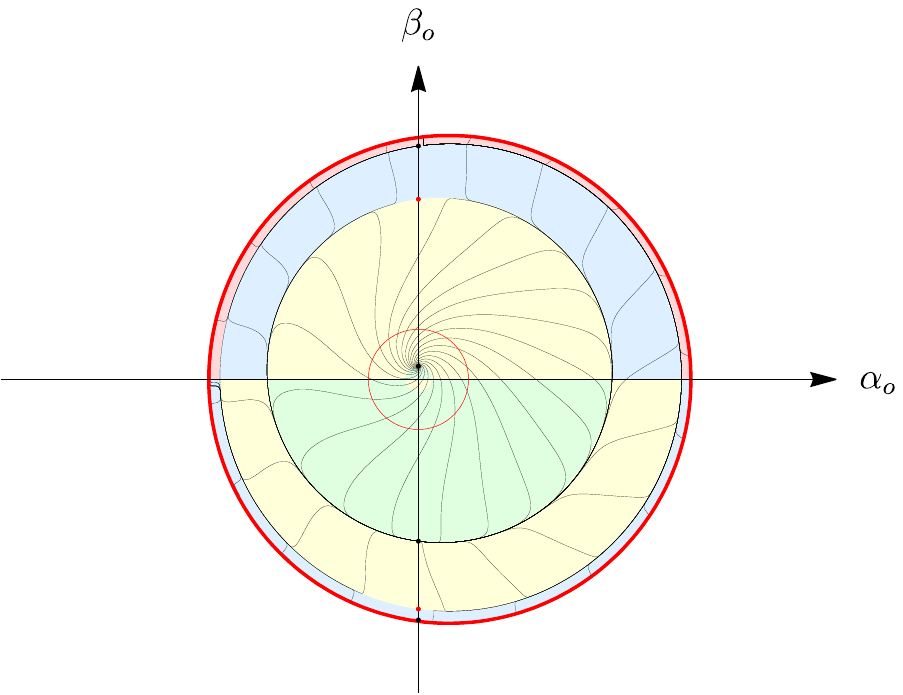}
	\caption{Prediction from conformal symmetry for the polarimetric image of the black hole at the center of M87 as seen at the EHT ($\theta_o=15^\circ$). Regions of different $m$ are colored according to the same conventions as in Fig.~\ref{fig:FaceOnEdgeOn}. Small dots along the vertical axis indicate the apparent locations of the poles.}
	\label{fig:M87}
\end{figure}

In Fig.~\ref{fig:M87}, we plot the corresponding polarimetric image for an observer situated $15^\circ$ off-axis, which is the angle relevant for observations of the black hole at the center of the galaxy M87. This generic image is considerably more complicated than the special cases considered in  Fig.~\ref{fig:FaceOnEdgeOn}, although it retains a (distorted) distinctive `whorl'-like pattern that could serve as a diagnostic for a high-spin black hole. 

An interferometer such as the EHT can reconstruct images of the linear polarization Stokes parameters $Q$ and $U$, or equivalently the complex polarization $P=Q+iU=mIe^{2i\chi}$, where $I$ is the Stokes intensity, $m$ is the degree of polarization, and $\chi$ denotes the Electric Vector Polarization Angle (EVPA) \cite{Roberts1994}. Our prediction \eqref{eq:Prediction} can be directly compared against the EVPA in reconstructed images via the relation $\tan{\chi}=-\mathcal{E}_\alpha/\mathcal{E}_\beta$.

Finally, we wish to reiterate that although the calculations in this paper were explicitly carried out only for the precisely extremal black hole with $a=M$, we expect our results to still hold (to leading order in $\epsilon$) for near-extremal black holes with a small deviation from extremality $a=M\sqrt{1-\epsilon^2}$ with $\epsilon\ll1$. For more details, the reader is invited to consult Sec.~IV in Ref.~\cite{Gralla2016b} and Sec.~4 in Ref.~\cite{Lupsasca2018}.

\section{Propagation of light and its polarization around a black hole}
\label{sec:GeometricOptics}

In this section, we review the standard treatment of the propagation of light and its polarization in the background of a rotating black hole \cite{Chandrasekhar1983}. We most closely follow the conventions of Ref.~\cite{Li2009}.

\subsection{Kerr geometry}

The Kerr metric describes astrophysically realistic rotating black holes of mass $M$ and angular momentum $J=aM$. In Boyer-Lindquist coordinates $\pa{t,r,\theta,\phi}$, the Kerr line element is
\begin{subequations}
\label{eq:Kerr}
\begin{gather}
	ds^2=-\frac{\Delta}{\Sigma}\pa{\ed t-a\sin^2{\theta}\ed\phi}^2+\frac{\Sigma}{\Delta}\ed r^2+\Sigma\ed\theta^2+\frac{\sin^2{\theta}}{\Sigma}\br{\pa{r^2+a^2}\ed\phi-a\ed t}^2,\\
	\Delta(r)=r^2-2Mr+a^2,\qquad
	\Sigma(r,\theta)=r^2+a^2\cos^2{\theta}.
\end{gather}
\end{subequations}
This metric admits two Killing vectors $\pd_t$ and $\pd_\phi$ generating time-translation symmetry and axisymmetry, respectively. In addition to these isometries, the Kerr metric also admits the existence of an irreducible symmetric Killing tensor\footnote{A Killing tensor satisfies $\nabla_{(\lambda}K_{\mu\nu)}=0$. The antisymmetric tensor $J_{\mu\nu}=-J_{\nu\mu}$ satisfies the Killing-Yano equation $\nabla_{(\lambda}J_{\mu)\nu}=0$.}
\begin{align}
	\label{eq:KerrKilling}
	K_{\mu\nu}=-{J_\mu}^\lambda J_{\lambda\nu},\qquad
	J=a\cos{\theta}\ed r\wedge\pa{\ed t-a\sin^2{\theta}\ed\phi}-r\sin{\theta}\ed\theta\wedge\br{a\ed t-\pa{r^2+a^2}\ed\phi}.
\end{align}

Many special properties of the Kerr geometry arise from the existence of a special complex null tetrad $\cu{l,n,m,\bar{m}}$, where $\cu{l,n}$ is a pair of real null vectors and $m$ is a complex null vector such that
\begin{subequations}
\label{eq:Tetrad}
\begin{align}
	l\cdot l=n\cdot n=m\cdot m=\bar{m}\cdot\bar{m}=0,\\
	l\cdot m=l\cdot\bar{m}=n\cdot m=n\cdot\bar{m}=0,\\
	l\cdot n=-1,\qquad
	m\cdot\bar{m}=1,\\
	\label{eq:MetricTetrad}
	g_{\mu\nu}=2\pa{m_{(\mu}\bar{m}_{\nu)}-l_{(\mu}n_{\nu)}}.
\end{align}
\end{subequations}
A possible choice of null tetrad obeying all these conditions in Kerr is\footnote{This choice is not unique: for instance, the conditions \eqref{eq:Tetrad} are invariant under rescalings $\pa{l,n}\to\pa{F^{-1}l,Fn}$ for any scalar function $F$. The standard choice for Kerr is the Kinnersley tetrad $\cu{l',n',m,\bar{m}}$, where $\pa{l',n'}=\pa{F^{-1}l,nF}$ with $F=\sqrt{\Delta/2\Sigma}$ (see, $e.g.$, Ref.~\cite{Chandrasekhar1983}). We have performed this rescaling in order to ensure that the tetrad \eqref{eq:KerrTetrad} has the simple near-horizon limit \eqref{eq:NHEKTetrad}.}
\begin{subequations}
\label{eq:KerrTetrad}
\begin{align}
	l&=\frac{1}{\sqrt{2\Delta\Sigma}}\br{\pa{r^2+a^2}\pd_t+\Delta\pd_r+a\pd_\phi},\\
	n&=\frac{1}{\sqrt{2\Delta\Sigma}}\br{\pa{r^2+a^2}\pd_t-\Delta\pd_r+a\pd_\phi},\\
	m&=\frac{1}{\sqrt{2}(r+ia\cos{\theta})}\pa{ia\sin{\theta}\pd_t+\pd_\theta+\frac{i}{\sin{\theta}}\pd_\phi}.
\end{align}
\end{subequations}

Under the geometric optics approximation, photons propagate along null geodesics. In the Kerr geometry \eqref{eq:Kerr}, this implies that a photon passing through a point $x^\mu$ in the spacetime has a four-momentum $p=p_\mu\ed x^\mu$ of the form
\begin{align}
	\label{eq:KerrGeodesic}
	p(x^\mu,\omega,\ell,k)=-\omega\ed t\pm_r\frac{\sqrt{\mathcal{R}(r)}}{\Delta(r)}\ed r\pm_\theta\sqrt{\Theta(\theta)}\ed\theta+\ell\ed\phi,
\end{align}
where the two choices of sign $\pm_r$ and $\pm_\theta$ determine the radial and polar directions of travel, respectively. Here, we also introduced radial and polar potentials 
\begin{subequations}
\begin{align}
	\mathcal{R}(r)&=\br{\omega\pa{r^2+a^2}-a\ell}^2-k\Delta(r),\\
	\Theta(\theta)&=k-\pa{\ell\csc{\theta}-a\omega\sin{\theta}}^2,
\end{align}
\end{subequations}
with $\pa{\omega,\ell,k}$ denoting the photon's energy, component of angular momentum parallel to the axis of symmetry, and Carter constant, respectively:\footnote{\label{fn:CarterIntegral}The Carter constant $k$ and Carter integral $Q=p_\theta^2-\cos^2{\theta}\pa{a^2p_t^2-p_\phi^2\csc^2{\theta}}$ are related by $k=Q+(\ell-a\omega)^2$.}
\begin{align}
	\label{eq:KerrConserved}
	\omega=-p_t,\qquad
	\ell=p_\phi,\qquad
	k=K^{\mu\nu}p_\mu p_\nu
	=p_\theta^2+2ap_tp_\phi+a^2p_t^2\sin^2{\theta}+p_\phi^2\csc^2{\theta}.
\end{align}
These quantities are conserved along the photon's trajectory, as is the Penrose-Walker constant\footnote{The last step follows from the identity \eqref{eq:MetricTetrad}. The conserved quantity $\kappa$ is associated with the existence of a conformal Killing spinor, which every Petrov D spacetime admits. For the Kerr spacetime, one can show that $\ab{\kappa}^2=\kappa_1^2+\kappa_2^2=k\pa{f\cdot f}$.}
\begin{subequations}
\label{eq:PenroseWalker}
\begin{align}
	\kappa&=2p^\mu f^\nu\pa{l_{[\mu}n_{\nu]}-m_{[\mu}\bar{m}_{\nu]}}\pa{r-ia\cos{\theta}}\\
	&=2\br{\pa{p\cdot l}\pa{f\cdot n}-\pa{p\cdot m}\pa{f\cdot\bar{m}}}\pa{r-ia\cos{\theta}},
\end{align}
\end{subequations}
provided that the vector $f$ is orthogonal to $p$ and parallel transported along it,
\begin{align}
	f\cdot p=0,\qquad
	p^\mu\nabla_\mu f^\nu=0.
\end{align}

When $f$ is a unit-norm spacelike vector, $f\cdot f=1$, we may interpret it as the photon's linear polarization. Its parallel transport is then determined by the conservation of the Penrose-Walker constant \eqref{eq:PenroseWalker}, which is typically rewritten as
\begin{subequations}
\label{eq:KerrPW}
\begin{gather}
	\kappa=(A-iB)(r-ia\cos{\theta})
	\equiv\kappa_1+i\kappa_2,\\
	A=(p\cdot l)(f\cdot n)-(p\cdot n)(f\cdot l),\qquad
	iB=(p\cdot m)(f\cdot\bar{m})-(p\cdot\bar{m})(f\cdot m).
\end{gather}
\end{subequations}

When a photon of momentum \eqref{eq:KerrGeodesic} impinges upon the screen of a distant observer at polar angle $\theta_o$, it appears at Cartesian coordinates $(\alpha_o,\beta_o)$ on the observer's screen given by \cite{Cunningham1972,Cunningham1973,Bardeen1973,Gralla2018}
\begin{align}
	\label{eq:ScreenCoordinates}
	\alpha_o=-\frac{\ell}{\omega\sin{\theta_o}},\qquad
	\beta_o=\pm_o\frac{\sqrt{\Theta(\theta_o)}}{\omega},
\end{align}
where the sign $\pm_o$ is equal to that of $p_\theta$ at the observer. The observed linear polarization of the photon (or equivalently, the direction of the electromagnetic wave's electric field) is given by a vector with Cartesian components on the screen
\begin{align}
	\label{eq:PolarizationAngle}
	\vec{\mathcal{E}}=\pa{\mathcal{E}_\alpha,\mathcal{E}_\beta}
	=\frac{1}{\omega\pa{\beta_o^2+\gamma_o^2}}\pa{\beta_o\kappa_2-\gamma_o\kappa_1,\beta_o\kappa_1+\gamma_o\kappa_2},
	\qquad\gamma_o=-(\alpha_o+a\sin{\theta_o}),
\end{align}
which has unit norm because $\kappa_1^2+\kappa_2^2=\omega^2\pa{\beta_o^2+\gamma_o^2}=k$. The overall sign is irrelevant.

A black hole casts a shadow on the observer's screen whose edge corresponds to light rays that asymptote to bound photon orbits. Provided $0<a<M$, the contour of the shadow is traced by the curve $\pa{\alpha\pa{\tau},\beta\pa{\tau}}$ where
\begin{subequations}
\label{eq:Shadow}
\begin{align}
	\alpha\pa{\tau}&=\frac{\tau^2\pa{\tau-3M}+a^2\pa{\tau+M}}{a\pa{\tau-M}\sin{\theta_o}},\qquad
	\tau\in\br{\tau_-,\tau_+},\qquad
	\tau_\pm\equiv2M\br{1+\cos\pa{\frac{2}{3}\arccos{\pm\frac{a}{M}}}},\\
	\beta\pa{\tau}&=\pm_o\sqrt{a^2\cos^2{\theta_o}-\frac{\tau^3\br{\tau\pa{\tau-3M}^2-4a^2M}+\br{\tau^2\pa{\tau-3M}+a^2\pa{\tau+M}}^2\cot^2{\theta_o}}{a^2\pa{\tau-M}^2}}.
\end{align}
\end{subequations}
The extremal regime $a\to M$ is subtle \cite{Gralla2018}. In that limit, the shadow develops a vertical edge at\footnote{This vertical edge is visible to observers in the range $\theta_o\in\br{\theta_c,\pi-\theta_c}$ but disappears above the critical angle $\theta_c=\arctan\pa{4/3}^{1/4}\approx47^\circ$, which is why it only appears in the depiction of the `edge-on' case in Figs.~\ref{fig:FaceOnEdgeOn} and \ref{fig:M87}. The reader is invited to consult Refs.~\cite{Gralla2018,Lupsasca2018} for a discussion of this so-called `NHEKline' and its features.}
\begin{align}
	\label{eq:NHEKline}
	\alpha=-2M\csc{\theta_o},\qquad
	\ab{\beta}<M\sqrt{3+\cos^2{\theta_o}-4\cot^2{\theta_o}},
\end{align}
with the remainder of the contour given by the curve
\begin{subequations}
\label{eq:ExtremalShadow}
\begin{align}
	M\alpha\pa{\tau}&=\pa{\tau^2-M^2-2M\tau}\csc{\theta_o},\qquad
	\tau\in(M,4M),\\
	M\beta\pa{\tau}&=\pm_o\sqrt{\tau^3\pa{4M-\tau}+M^4\cos^2{\theta_o}-\pa{\tau^2-M^2-2M\tau}^2\cot^2{\theta_o}}.
\end{align}
\end{subequations}

\subsection{NHEK geometry}

When the angular momentum of a rotating black hole nears saturation of the Kerr bound $\ab{J}\le M^2$, the region of spacetime in the vicinity of its event horizon develops a throat of divergent proper depth \cite{Bardeen1972}. To resolve physics in this near-horizon region, one can introduce Bardeen-Horowitz coordinates
\begin{align}
	\label{eq:NHEKlimit}
	t=\frac{2M}{\epsilon^p}T,\qquad
	r=M(1+\epsilon^pR),\qquad
	\phi=\Phi+\frac{T}{\epsilon^p},
\end{align}
and then zoom into the horizon by taking the scaling limit $\epsilon\to0$, with either
\begin{subequations}
\begin{align}
	\label{eq:ExtremeLimit}
	&\text{Extremal limit}:&&
	a=M,&&
	p=1,\\
	\label{eq:NearExtremeLimit}
	&\text{Near-extremal limit}:&&
	a=M\sqrt{1-\epsilon^2},&&
	0<p<1.
\end{align}
\end{subequations}
Applying this procedure to the Kerr metric \eqref{eq:Kerr} produces the Near-Horizon Extreme Kerr (NHEK) line element \cite{Bardeen1999}
\begin{subequations}
\label{eq:NHEK}
\begin{gather}
	d\hat{s}^2=2M^2\Gamma\br{-R^2\ed T^2+\frac{\ed R^2}{R^2}+\ed\theta^2+\Lambda^2\pa{\ed\Phi+R\ed T}^2},\\
	\Gamma(\theta)=\frac{1+\cos^2{\theta}}{2},\qquad
	\Lambda(\theta)=\frac{2\sin{\theta}}{1+\cos^2{\theta}},
\end{gather}
\end{subequations}
which forms a spacetime solution to Einstein's equations in its own right. From now on, we denote contractions with respect to this hatted metric by $\circ$ and reserve $\cdot$ for contractions in Kerr.

In the remainder of this paper, we will study electromagnetic emissions from the near-horizon region described by the throat geometry \eqref{eq:NHEK}. For simplicity, we will restrict our attention to the case of a precisely extremal black hole with $a=M$ and use the scaling limit \eqref{eq:ExtremeLimit}. However, we expect the polarimetric images we obtained to be identical to leading order in $\epsilon$ in the scaling limit \eqref{eq:NearExtremeLimit} relevant for the physically realistic case of a near-extremal black hole such as M87*. Moreover, as subleading corrections are power-suppressed in $\epsilon$, we expect this leading-order result to be an excellent approximation for M87* ($\epsilon<0.2$).

The NHEK geometry has an enlarged $\mathsf{SL}(2,\mathbb{R})\times\mathsf{U}(1)$ isometry group generated by the four Killing vector fields
\begin{align}
	W_0=\pd_\Phi,\qquad
	H_+=\pd_T,\qquad
	H_0=T\pd_T-R\pd_R,\qquad
	H_-=\pa{T^2+\frac{1}{R^2}}\pd_T-2TR\pd_R-\frac{2}{R}\pd_\Phi,
\end{align}
thanks to an enhancement of the time translation symmetry $H_+$ to an $\mathsf{SL}(2,\mathbb{R})$ global conformal symmetry that also includes dilations $H_0$ and special conformal transformations $H_-$.

In the near-horizon limit, the irreducible Killing tensor \eqref{eq:KerrKilling} in Kerr becomes a reducible Killing tensor in NHEK that is given (up to a mass term) by the Casimir of $\mathsf{SL}(2,\mathbb{R})\times\mathsf{U}(1)$,
\begin{align}
	\hat{K}^{\mu\nu}=M^2\hat{g}^{\mu\nu}-H_0^\mu H_0^\nu+\frac{1}{2}\pa{H_+^\mu H_-^\nu+H_-^\mu H_+^\nu}+W_0^\mu W_0^\nu.
\end{align}
Meanwhile, the extreme Kerr tetrad \eqref{eq:KerrTetrad} descends to the NHEK tetrad
\begin{subequations}
\label{eq:NHEKTetrad}
\begin{align}
	\hat{l}&=\frac{1}{2M\sqrt{\Gamma}}\pa{\frac{1}{R}\pd_T+R\pd_R-\pd_\Phi},\\
	\hat{n}&=\frac{1}{2M\sqrt{\Gamma}}\pa{\frac{1}{R}\pd_T-R\pd_R-\pd_\Phi},\\
	\hat{m}&=\frac{1}{\sqrt{2}M}\pa{\frac{1}{1+i\cos{\theta}}\pd_\theta+\frac{\cos{\theta}+i}{2\sin{\theta}}\pd_\Phi},
\end{align}
\end{subequations}
which also satisfies the null tetrad conditions \eqref{eq:Tetrad} with respect to the NHEK metric \eqref{eq:NHEK}.

Under the geometric optics approximation, a photon passing through a point $X^\mu$ in the NHEK geometry \eqref{eq:NHEK} has a four-momentum $P=P_\mu\ed X^\mu$ of the form
\begin{align}
	\label{eq:NHEKGeodesic}
	P(X^\mu,E,L,K)=-E\ed T\pm_r\frac{\sqrt{\pa{E+LR}^2-KR^2}}{R^2}\ed R\pm_\theta\sqrt{K-\frac{L^2}{\Lambda^2}}\ed\theta+L\ed\Phi,
\end{align}
where the signs determine the direction of travel, and $\pa{E,L,K}$ denote the photon's near-horizon energy (with respect to NHEK time), component of angular momentum parallel to the axis of symmetry, and Carter constant, respectively:\footnote{Observe that $K-L^2=-h_0^2+\frac{1}{2}\pa{h_+h_-+h_-h_+}$, where $h_i=H_i^\mu u_\mu$ are the conserved quantities associated with the generators of $\mathsf{SL}(2,\mathbb{R})$. Since these are not independent of each other, we use the $\mathsf{SL}(2,\mathbb{R})$ Casimir $K-L^2$, which is in involution with all the $h_i$ \cite{AlZahrani2011}. This is exactly analogous to exploiting the conservation of $J_z$ and $J^2$, rather than $(J_x,J_y,J_z)$, in a problem with $\mathsf{SO}(3)$ symmetry.}
\begin{align}
	\label{eq:NHEKConserved}
	E=-P_T,\qquad
	L=P_\Phi,\qquad
	K=\hat{K}^{\mu\nu}P_\mu P_\nu
	=\pa{1-\frac{1}{2\Gamma}}\br{\pa{P_\Phi-\frac{P_T}{R}}^2-P_R^2R^2}+\frac{1}{2\Gamma}\pa{P_\Theta^2+\frac{P_\Phi^2}{\Lambda^2}}.
\end{align}
These quantities are conserved along the photon's trajectory, as is the Penrose-Walker constant\footnote{Note that $M(1-i\cos{\theta})$ is the near-horizon limit \eqref{eq:NHEKlimit} of its Kerr analogue $r-ia\cos{\theta}$. Up to some factors of $M$, these are the Weyl tensor components $\Psi_2$ (in the Newman-Penrose formalism) of NHEK and Kerr, respectively.}
\begin{subequations}
\label{eq:NHEKPW}
\begin{gather}
	\mathcal{K}=M\pa{\mathcal{A}-i\mathcal{B}}(1-i\cos{\theta}),\\
	\mathcal{A}=(P\circ\hat{l})(F\circ\hat{n})-(P\circ\hat{n})(F\circ\hat{l}),\qquad
	i\mathcal{B}=(P\circ\hat{m})(F\circ\bar{\hat{m}})-(P\circ\bar{\hat{m}})(F\circ\hat{m}),
\end{gather}
\end{subequations}
provided that the vector $F$ is orthogonal to $P$ and parallel transported along it,
\begin{align}
	F\circ P=0,\qquad
	P^\mu\hat{\nabla}_\mu F^\nu=0.
\end{align}
As in Kerr, when the vector $F$ is a unit-norm spacelike vector, $F\circ F=1$, we may interpret it as the photon's linear polarization. Its parallel transport is then determined by the conservation of the Penrose-Walker constant \eqref{eq:NHEKPW}.

\section{Near-horizon electromagnetic emissions from a high-spin black hole}
\label{sec:Computation}

In the geometric optics approximation, a thin beam of photons is modeled by a narrow bundle of rays. Together, these rays form a null geodesic congruence. If we give each photon in the beam the same energy, then we can uniquely fix the vector field tangent to the congruence by setting it equal, at every point $x^\mu$ within the beam, to the momentum $p(x^\mu)$ of the photon passing through that point. For a light beam in Kerr, this results in the vector field \eqref{eq:KerrGeodesic}.

Additionally, we can also define a polarization vector field $f(x^\mu)$ at every point within the light beam. This spacelike vector field must have unit norm and be everywhere orthogonal to $p(x^\mu)$, the generator of the null congruence. Thus, a beam of radiation, which is completely characterized by its intensity and polarization, can be described by two vector fields $p$ and $f$ obeying the null geodesic and parallel transport equations\footnote{In general, $f$ is a complex vector field subject to the normalization condition $f\cdot\bar{f}=1$. Here, we assume the light is linearly polarized light, in which case $f$ is real and $f\cdot f=1$.}
\begin{align}
	\label{eq:Beam}
	p^\mu\nabla_\mu p^\nu=p^\mu\nabla_\mu f^\nu=0,\qquad
	p\cdot p=p\cdot f=0,\qquad
	f\cdot f=1.
\end{align}

Next, suppose that we shine such a beam of radiation out of (or into) the near-horizon region of an extreme Kerr black hole. The NHEK portion of the beam is then described by a momentum $P(X^\mu)$ and polarization $F(X^\mu)$ obeying the NHEK analogues of Eqs.~\eqref{eq:Beam},
\begin{align}
	\label{eq:NHEKBeam}
	P^\mu\hat{\nabla}_\mu P^\nu=P^\mu\hat{\nabla}_\mu F^\nu=0,\qquad
	P\circ P=P\circ F=0,\qquad
	F\circ F=1.
\end{align}

However, the converse is not true: not all NHEK vector fields $P(X^\mu)$ and $F(X^\mu)$ solving these equations describe beams of radiation that escape the NHEK region and connect to a distant observer. Those that do have to obey an additional constraint: they must arise as the near-horizon limits of corresponding Kerr vector fields $p(x^\mu)$ and $f(x^\mu)$. As we will now show, this requirement imposes additional symmetry constraints on $P(X^\mu)$ and $F(X^\mu)$. This is a specific instance of a more general phenomenon: the near-horizon limit of a field in extreme Kerr acquires a definite weight under NHEK dilations by the scaling parameter $\epsilon$ \cite{Gralla2016b}. Further details are presented in App.~\ref{app:Expansion}.

\subsection{Generic photons}

Rotating black holes exhibit a surprising phenomenon known as superradiance: they can amplify the energy of an object scattered off the horizon, as long as its incident energy does not exceed the so-called superradiant bound
\begin{align}
	\omega-\Omega_H\ell\le0,
\end{align}
where $\Omega_H$ denotes the angular frequency of the event horizon. For an extremal black hole, $\Omega_H=1/(2M)$, and we refer to photons not saturating the superradiant bound ($\omega\neq\Omega_H\ell$) as \textit{generic}.

In the near-horizon limit \eqref{eq:NHEKlimit}, the Kerr four-momentum \eqref{eq:KerrGeodesic} of a generic photon entering the NHEK region is
\begin{subequations}
\label{eq:H1}
\begin{align}
	p(x^\mu,\omega,\ell,k)&=-\frac{2M\omega-\ell}{\epsilon}\ed\pa{T\pm_r\frac{1}{R}}+\O{\epsilon^0}\\
	&=P\pa{X^\mu,\frac{2M\omega-\ell}{\epsilon},0,0}+\O{\epsilon^0},
\end{align}
\end{subequations}
where the second line follows by comparison with the four-momentum \eqref{eq:NHEKGeodesic} of a NHEK photon. Hence, the conserved quantities $(\omega,\ell,k)$ along the photon's trajectory in Kerr are related to their NHEK analogues $(E,L,K)$ by\footnote{This relation reflects the high energy cost incurred by near-horizon photons that climb out of the throat: an outgoing photon undergoes a parametrically large redshift as it escapes from the parametrically deep gravitational well of the black hole. (`Parametrically large/small' means `diverging/vanishing' as $\epsilon\to0$.) Conversely, from the perspective of an observer lying at a Boyer-Lindquist radius of order $\epsilon$ from the horizon, an infalling particle with $\omega-\Omega_H\ell\neq0$ in Kerr must necessarily have a NHEK energy that diverges as the observer is pushed deeper into the throat ($E\to\infty$ as $\epsilon\to0$). For this reason, the throat is naturally a site of high-energy collisions that can supply the highly energetic photons needed for our signal \cite{Banados2009,Gralla2016b}.}
\begin{align}
	\label{eq:ConservedH1}
	\omega-\Omega_H\ell=\frac{\epsilon E}{2M},\qquad
	L=0,\qquad
	K=0,
\end{align}
in accordance with Eqs.~\eqref{eq:GeneralH} with $H=1$. Moreover, note that under a NHEK dilation $(T,R)\to(T',R')=(T/\epsilon,\epsilon R)$,
\begin{align}
	P(X^\mu,E,0,0)=-E\ed\pa{T\pm_r\frac{1}{R}}\quad\longrightarrow\quad
	P(X'^\mu,E,0,0)=-\frac{E}{\epsilon}\ed\pa{T\pm_r\frac{1}{R}}
	=\frac{1}{\epsilon}P(X^\mu,E,0,0).
\end{align}
Infinitesimally, this imposes a symmetry condition on a narrow bundle of rays surrounding the photon,
\begin{align}
	\L_{H_0}P(X^\mu,E,0,0)=P(X^\mu,E,0,0),
\end{align}
in agreement with Eq.~\eqref{eq:Weights} with $H=1$. This proves that a generic beam of photons in Kerr enters (or leaves) the near-horizon region along a null geodesic congruence with the largest allowed weight. (We show that $H\le1$ in App.~\ref{app:Expansion}.)

In fact, we can make a stronger statement: any generic beam of photons entering (or leaving) the near-horizon region must do so along a principal null congruence (PNC) of the NHEK geometry. There are two such congruences: the ingoing one generated by $\hat{n}$ and the outgoing one generated by $\hat{l}$, where $(\hat{l},\hat{n})$ are the real vectors in the NHEK tetrad \eqref{eq:NHEKTetrad}. Indeed, note that
\begin{subequations}
\label{eq:OutgoingMomentum}
\begin{align}
	&\text{Outgoing PNC }(\pm_r=1):&&
	P(X^\mu,E,0,0)=-E\ed\pa{T+\frac{1}{R}}=\frac{E}{MR\sqrt{\Gamma}}\hat{l},\\
	&\text{Ingoing PNC }(\pm_r=-1):&&
	P(X^\mu,E,0,0)=-E\ed\pa{T-\frac{1}{R}}=\frac{E}{MR\sqrt{\Gamma}}\hat{n}.
\end{align}
\end{subequations}

\subsection{Photons at the superradiant bound}

Notice that if we fine-tune the Kerr photon to be at the superradiant bound $\omega=\Omega_H\ell$, then $2M\omega-\ell=0$ and as such, the leading term in the expansion \eqref{eq:H1} vanishes.\footnote{The limits $\ell\to 2M\omega$ and $\epsilon\to0$ do not commute. In order to ensure that the superradiant photons solve the NHEK geodesic equation, the limit $\ell\to 2M\omega$ must be taken first.} Hence, the expansion starts at the next order, with
\begin{subequations}
\label{eq:H0}
\begin{align}
	p(x^\mu,\omega,\ell,k)&=\pm_r\frac{\sqrt{\ell^2-k}}{R}\ed R\pm_\theta\sqrt{k-\frac{\ell^2}{\Lambda^2}}\ed\theta+\ell\ed\Phi+\O{\epsilon}\\
	&=P(X^\mu,0,\ell,k)+\O{\epsilon},
\end{align}
\end{subequations}
where the second line follows by comparison with the four-momentum \eqref{eq:NHEKGeodesic} of a NHEK photon. Hence, the conserved quantities $(\omega,\ell,k)$ along the photon's trajectory in Kerr are related to their NHEK analogues $(E,L,K)$ by
\begin{align}
	\label{eq:ConservedH0}
	\omega-\Omega_H\ell=\frac{E}{2M}=\O{\epsilon},\qquad
	\ell=L,\qquad
	k=K,
\end{align}
in accordance with Eqs.~\eqref{eq:GeneralH} with $H=0$. Moreover, note that under a NHEK dilation $(T,R)\to(T',R')=(T/\epsilon,\epsilon R)$,
\begin{align}
	P(X^\mu,0,L,K)=\pm_r\frac{\sqrt{L^2-K}}{R}\ed R\pm_\theta\sqrt{K-\frac{L^2}{\Lambda^2}}\ed\theta+L\ed\Phi\quad\longrightarrow\quad
	P(X'^\mu,0,L,K)=P(X^\mu,0,L,K).
\end{align}
Infinitesimally, this imposes a symmetry condition on a narrow bundle of rays surrounding the photon,
\begin{align}
	\L_{H_0}P(X^\mu,0,L,K)=0,
\end{align}
in agreement with Eq.~\eqref{eq:Weights} with $H=0$. Although such null geodesic congruences may appear mathematically fine-tuned, they are abundantly realized in nature as the photon beams emitted by an orbiting hot spot or the near-horizon portion of a slowly accreting disk \cite{Gralla2018,Lupsasca2018}.

\subsection{Symmetric source polarization}
\label{sec:SourcePolarization}

We now imagine that every point in the near-horizon region of an extreme Kerr black hole emits beams of radiation in every direction with some (possibly vanishing) intensity. Thus, at every point $X^\mu$, we could have photons emitted with any momentum $P(X^\mu,E,L,K)$ and corresponding polarization $F(X^\mu,E,L,K)$. From the preceding discussion, we know that such light beams may escape to the asymptotically flat region with two qualitatively different behaviors:
\begin{enumerate}
\item
Photons in the beam have NHEK four-momentum $P(X^\mu,E/\epsilon,0,0)$ aligned with the outgoing PNC [Eq.~\eqref{eq:OutgoingMomentum}]. They have vanishing NHEK angular momentum and Carter constant ($L=K=0$), but parametrically large NHEK energy ($E\sim1/\epsilon$). They reach the far region as generic $H=1$ photons with finite energy above extremality $\omega-\Omega_H\ell=\epsilon E/2M\neq0$ and arbitrary Kerr Carter constant $k$ [Eqs.~\eqref{eq:ConservedH1}]. They can appear anywhere within the shadow of the black hole.
\item
Photons in the beam have NHEK four-momentum $P(X^\mu,0,L,K)$ [Eq.~\eqref{eq:H0}] and vanishing NHEK energy ($E=0$). They reach the far region as $H=0$ photons at the superradiant bound $\omega=\Omega_H\ell$ with Kerr angular momentum and Carter constant given by $\ell=L$ and $k=K$ [Eqs.~\eqref{eq:ConservedH0}].\footnote{This is true for $H=0$ photons. We neglect the measure-zero set of photons with $H<0$, which may only come out with $\omega=\ell=k=0$.} They can only appear on the vertical edge of the black hole's shadow (the `NHEKline').
\end{enumerate}

In both cases, the near-horizon polarization must necessarily obey Eqs.~\eqref{eq:NHEKBeam} and \eqref{eq:Weights},
\begin{align}
	\label{eq:NHEKPolarization}
	P^\mu\hat{\nabla}_\mu F^\nu=0,\qquad
	P\circ F=0,\qquad
	F\circ F=1,\qquad
	\L_{H_0}F=0.
\end{align}
This last dilation-invariance condition is derived in App.~\eqref{app:Expansion} from properties of the near-horizon limit \eqref{eq:NHEKlimit}. It is one of the additional symmetry constraints alluded to below Eqs.~\eqref{eq:NHEKBeam}.
 
A spacetime tends to impart its symmetries to the physics it drives. For instance, since accretion onto a black hole is a gravitational process, we expect accretion disks to inherit the stationarity and axisymmetry of the Kerr metric \eqref{eq:Kerr}---this standard assumption is largely borne out by numerical simulations (at least in the slow accretion regime). By this reasoning, we may also expect that the physical processes responsible for electromagnetic emissions from NHEK (whatever they are) respect the symmetries of the spacetime. Thus, we make the natural assumption that
\begin{align}
	\label{eq:SymmetryConditions}
	\L_{W_0}F=\L_{H_\pm}F=0.
\end{align}
That is, we suppose that the near-horizon source configuration produces radiation possessing the same symmetries as the NHEK geometry \eqref{eq:NHEK}: stationarity, axisymmetry, dilation-invariance, and special-conformal symmetry. This will be our only physical assumption and it will lead us to a clear signature. It would be interesting to find an example of a physical mechanism that would produce such a polarization profile at the source---in the interim, we simply note that traditional simulations also appear to predict a whorl-like pattern in certain regimes \cite{Gold2017,Akiyama2017,Moscibrodzka2017,JimenezRosales2018}.

For $H=1$ photons with four-momentum \eqref{eq:H1}, we find that there is a unique choice (up to an irrelevant overall sign) of source polarization in NHEK obeying Eqs.~\eqref{eq:NHEKPolarization} together with the symmetry-based assumption \eqref{eq:SymmetryConditions}:
\begin{align}
	\label{eq:Polarization}
	F=\frac{1}{2M\Gamma}\pa{\mp_r\pd_\theta+\frac{\cos{\theta}}{\Lambda}\pd_\Phi}
	=\frac{1}{\sqrt{2}}\pa{\frac{\cos{\theta}\mp_ri}{\cos{\theta}+i}\hat{m}+\frac{\cos{\theta}\pm_ri}{\cos{\theta}-i}\hat{\bar{m}}}.
\end{align}
Plugging this into Eqs.~\eqref{eq:NHEKPW}, one immediately sees that $\mathcal{A}=\mathcal{B}=0$ because $F\circ\hat{l}=F\circ\hat{n}=0$ and $P\circ\hat{m}=P\circ\hat{\bar{m}}=0$ by Eqs.~\eqref{eq:Tetrad} and \eqref{eq:OutgoingMomentum}. As such, we obtain
\begin{align}
	\label{eq:NaivePolarization}
	\mathcal{K}=0.
\end{align}

However, this result is misleading because it does not imply the vanishing of the Penrose-Walker constant $\kappa$ along the beam in Kerr. For $H=1$ geodesics (see App.~\ref{app:Subleading}),
\begin{align}
	\kappa=\frac{\hat{\kappa}}{\epsilon}+\tilde{\kappa}+\O{\epsilon},\qquad
	\hat{\kappa}=\mathcal{K}.
\end{align}
Although we have found that the leading-order $\O{\epsilon^{-1}}$ piece of $\kappa$ vanishes, its subleading $\O{\epsilon^0}$ piece does not: we show in App.~\ref{app:Subleading} that to leading-order in $\epsilon$,
\begin{align}
	\label{eq:LeadingKappa}
	\kappa=\tilde{\kappa}=\omega\pa{\beta_s\pm_ri\gamma_s},
\end{align}
where, by analogy with Eqs.~\eqref{eq:ScreenCoordinates}--\eqref{eq:PolarizationAngle}, we defined `source' quantities
\begin{align}
	\alpha_s=-\frac{\ell}{\omega\sin{\theta_s}},\qquad
	\beta_s=\pm_s\frac{\sqrt{\Theta(\theta_s)}}{\omega},\qquad
	\gamma_s=-(\alpha_s+M\sin{\theta_s}).
\end{align}
Here, the choice of sign $\pm_s$ corresponds to the initial direction of the photon's polar motion. 

We conclude that under our symmetry-based assumption \eqref{eq:SymmetryConditions}, this formula for $\kappa$ holds for all the photons received by a distant observer from the near-horizon region of an extreme Kerr black hole.

\subsection{Polarization at the observer}

From Eq.~\eqref{eq:LeadingKappa}, we can now read off
\begin{align}
	\kappa_1=\omega\beta_s,\qquad
	\kappa_2=\pm_r\omega\gamma_s.
\end{align}
Finally, plugging this into Eq.~\eqref{eq:PolarizationAngle} yields the polarization profile measured by a distant observer:
\begin{align}
	\label{eq:ObservedPolarization}
	\vec{\mathcal{E}}=\frac{1}{\pa{\beta_o^2+\gamma_o^2}}\pa{\pm_r\beta_o\gamma_s-\gamma_o\beta_s,\beta_o\beta_s\pm_r\gamma_o\gamma_s}.
\end{align}
This quantity depends on both the observer parameters $\pa{\theta_o,\alpha_o,\beta_o}$ and the source parameters $\pa{\theta_s,\alpha_s,\beta_s}$, as well as the discrete sign choices $\pa{\pm_r,\pm_s,\pm_o}$. We will now eliminate the variables $\pa{\theta_s,\alpha_s,\beta_s,\pm_r,\pm_s,\pm_o}$ in favor of $\pa{\theta_o,\alpha_o,\beta_o}$, which are the only variables parameterizing observations.

First, we have $\pm_o=\sign(\beta_o)$ by definition. Next, since we are interested in photons that are eventually outgoing, we set $\pm_r=1$. (This choice does not exclude initially ingoing photons that reach a radial turning point before escaping to the far region and is therefore still completely general---see App.~\ref{app:GeodesicEquation} for details.) Since the conserved quantities of the geodesics are the same at the observer and the source, we can eliminate $(\alpha_s,\beta_s)$ in favor of $(\alpha_o,\beta_o)$:
\begin{align}
	\label{eq:SourceScreen}
	\alpha_s=-\frac{\ell}{\omega\sin{\theta_s}}
	=\frac{\sin{\theta_o}}{\sin{\theta_s}}\alpha_o,\qquad
	\beta_s=\pm_s\frac{\sqrt{\Theta(\theta_s)}}{\omega}
	=\pm_s\sqrt{\beta_o^2+\gamma_o^2-\gamma_s^2},
\end{align}
where the last equality follows from the conservation of $k/\omega^2=\beta_i^2+\gamma_i^2$ for $i\in\cu{o,s}$. Finally, we fix $\theta_s$ and $\pm_s$ by solving the geodesic equation, as described in App.~\ref{app:GeodesicEquation}. Putting everything together results in our prediction \eqref{eq:Prediction}.

\section*{Acknowledgments}

This work was supported in part by NSF grant 1205550 to Harvard University. DK gratefully acknowledges support from DOE grant DE-SC0009988. DG gratefully acknowledges support from NSF GRFP grant DGE1144152. The authors thank Geoffrey Comp\`ere, Sheperd Doeleman, Samuel Gralla, Michael D. Johnson, and Achilleas Porfyriadis for fruitful conversations. Many of these discussions took place at the Black Hole Initiative at Harvard University, which is supported by a grant from the John Templeton Foundation.

\appendix

\section{Geodesics connecting the near-horizon and asymptotically flat regions}
\label{app:Expansion}

It is natural to ask what happens to the Kerr momentum $p$ and polarization $f$ of a light beam entering (or leaving) the NHEK region \eqref{eq:NHEK}. To answer this question, we first take the near-horizon limit \eqref{eq:NHEKlimit} of $p$ and $f$, resulting in
\begin{align}
	\label{eq:NearHorizonExpansions}
	p(x^\mu)=\epsilon^{-H}\br{\hat{p}(X^\mu)+\O{\epsilon}},\qquad
	f(x^\mu)=\epsilon^{-H'}\br{\hat{f}(X^\mu)+\O{\epsilon}},
\end{align}
for some integers $H$ and $H'$.\footnote{As discussed in Sec.~IV of Ref.~\cite{Gralla2016b}, these scaling powers may become fractional powers in the context of a near-extremal black hole.} We can then proceed along the same lines as Ref.~\cite{Gralla2016b}. We begin by noting, based on physical grounds, that neither $p$ nor $f$ is allowed to blow up on the event horizon. Demanding their regularity on the horizon requires that $H,H'\le1$. Next, plugging the leading-order near-horizon expansions \eqref{eq:NearHorizonExpansions} into Eqs.~\eqref{eq:Beam} yields
\begin{align}
	\label{eq:LeadingBeam}
	\hat{p}^\mu\hat{\nabla}_\mu\hat{p}^\nu=\hat{p}^\mu\hat{\nabla}_\mu\hat{f}^\nu=0,\qquad
	\hat{p}\circ\hat{p}=\hat{p}\circ\hat{f}=0,\qquad
	\epsilon^{-2H'}\hat{f}\circ\hat{f}=1,
\end{align}
where all the contractions and covariant derivatives appearing in these leading-order expressions are meant to be taken with respect to the NHEK metric \eqref{eq:NHEK}, rather than the Kerr metric \eqref{eq:Kerr} that is meant to be used in Eqs.~\eqref{eq:Beam}.

These equations indicate that, for any allowed weight $H\le1$, the NHEK vector field $\hat{p}$ describing the leading-order near-horizon beam is also tangent to a null geodesic congruence, like its Kerr counterpart $p$. This was to be expected: at every NHEK coordinate $X^\mu$ within the beam, $\hat{p}(X^\mu)$ must by definition equal the momentum of the photon passing through $X^\mu$. However, it is important to note that the Kerr conserved quantities \eqref{eq:KerrConserved} of the photons in the beam need not equal their NHEK counterparts \eqref{eq:NHEKConserved}---in general, their precise relation is given by
\begin{align}
	\label{eq:GeneralH}
	\omega-\Omega_H\ell=\epsilon^{-H+1}\frac{\hat{E}}{2M}
	=\frac{\epsilon E}{2M},\qquad
	\ell=\epsilon^{-H}\hat{L}
	=L,\qquad
	k=\epsilon^{-2H}\hat{K}
	=K,
\end{align}
where $\Omega_H=1/(2M)$ denotes the angular velocity of the horizon at extremality and the hatted conserved quantities are defined relative to $\hat{p}$ (so as to always be finite) while the unhatted quantities are defined relative to $P=\epsilon^{-H}\hat{p}$ (and therefore may be parametrically large or small). These nontrivial relations follow directly from the infrared limit \eqref{eq:NHEKlimit} and the near-horizon expansions \eqref{eq:NearHorizonExpansions}, with the various factors of $\epsilon$ arising from the scaling to a region of infinite redshift near the horizon.\footnote{These relations also appear in App.~B of Ref.~\cite{Gralla2016b}. As an example, note from the definitions \eqref{eq:KerrConserved} and \eqref{eq:NHEKConserved} that $\ell=p\cdot\pd_\phi$ and $\hat{L}=\hat{p}\circ\pd_\Phi$, where $\phi$ and $\Phi$ are the Kerr and NHEK angular coordinates, respectively. One can check that $\pd_\phi=\pd_\Phi+\O{\epsilon}$, so to leading order in $\epsilon$,
\begin{align*}
	\ell=p\cdot\pd_\phi
	=\epsilon^{-H}\br{\hat{p}+\O{\epsilon^1}}\circ\pd_\Phi
	=\epsilon^{-H}\br{\hat{L}+\O{\epsilon^0}}.
\end{align*}
}

On the other hand, the near-horizon limit of the polarization vector field is much more constrained by Eqs.~\eqref{eq:LeadingBeam}, which are manifestly inconsistent unless $H'=0$: indeed, if $H'\neq0$, then the norm of the near-horizon polarization vector $\hat{f}$ either vanishes or diverges. Both situations are unphysical: the latter because it is meaningless and the former because it requires the polarization to become longitudinal, or pure gauge.\footnote{Since the physical photon polarization is invariant under longitudinal shifts $f\to f+cp$, we must in general relax the parallel transport condition $p^\mu\nabla_\mu f^\nu=0$ to the weaker condition $p^\mu\nabla_\mu f^\nu\propto p^\nu$, which still guarantees the parallel transport of the orthogonality condition $p\cdot f=0$. In general, $f$ will have a longitudinal component with weight $H'=1$, but it can always be removed by a gauge transformation.}

The values of the weights $H$ and $H'$ are significant because they determine the leading behavior of the fields $p$ and $f$ under scale transformations of the near-horizon region: the Lie derivatives of $\hat{p}$ and $\hat{f}$ along the generator $H_0$ of NHEK dilations $(T,R)\to(T/\epsilon,\epsilon R)$ are fixed to be
\begin{align}
	\label{eq:Weights}
	\L_{H_0}\hat{p}=H\hat{p},\qquad
	\L_{H_0}\hat{f}=H'\hat{f}=0.
\end{align}

\section{Subleading order expansion}
\label{app:Subleading}

In this appendix, we extend the preceding discussion beyond leading order in the NHEK limit \eqref{eq:NHEKlimit}. Working to subleading order in $\epsilon$ (and recalling from App.~\ref{app:Expansion} that $H'=0$), the near-horizon expansions \eqref{eq:NearHorizonExpansions} of the momentum $p$ and polarization $f$ of a light beam in Kerr are extended to
\begin{align}
	p(x^\mu)=\epsilon^{-H}\br{\hat{p}(X^\mu)+\epsilon\tilde{p}(X^\mu)+\O{\epsilon^2}},\qquad
	f(x^\mu)=\hat{f}(X^\mu)+\epsilon\tilde{f}(X^\mu)+\O{\epsilon^2},
\end{align}
where the limit requires that the following symmetry properties be satisfied:
\begin{align}
	\label{eq:SubleadingSymmetryConditions}
	\L_{H_0}\hat{p}=H\hat{p},\qquad
	\L_{H_0}\hat{f}=0,\qquad
	\L_{H_0}\tilde{p}=(H-1)\tilde{p},\qquad
	\L_{H_0}\tilde{f}=-\tilde{f}.
\end{align}
Likewise, in the NHEK limit \eqref{eq:NHEKlimit}, the Penrose-Walker constant \eqref{eq:KerrPW} in Kerr has an expansion
\begin{align}
	\kappa=\epsilon^{-H}\br{\hat{\kappa}+\epsilon\tilde{\kappa}+\O{\epsilon^2}}.
\end{align}

We now focus on the case $H=1$ of relevance to the discussion in Sec.~\ref{sec:SourcePolarization} of generic photons, with leading momentum $\hat{p}$ as given in Eqs.~\eqref{eq:OutgoingMomentum},
\begin{align}
	\hat{p}=\frac{2M\omega-\ell}{2M^2\Gamma}\pa{\frac{1}{R^2}\pd_T\pm_r\pd_R-\frac{1}{R}\pd_\Phi}.
\end{align}
There is a unique corresponding choice \eqref{eq:Polarization} of leading-order polarization obeying the symmetry-based conditions \eqref{eq:SymmetryConditions},
\begin{align}
	\hat{f}=\frac{1}{\sqrt{2}}\pa{\frac{\cos{\theta}\mp_ri}{\cos{\theta}+i}\hat{m}+\frac{\cos{\theta}\pm_ri}{\cos{\theta}-i}\hat{\bar{m}}},
\end{align}
which solves Eqs.~\eqref{eq:Beam} to leading order in $\epsilon$. As claimed in Eq.~\eqref{eq:NaivePolarization}, with these choices, we find that the Penrose-Walker constant vanishes to leading order,
\begin{align}
	\hat{\kappa}=0.
\end{align}
Next, to subleading order in the near-horizon limit \eqref{eq:NHEKlimit}, the Kerr four-momentum \eqref{eq:KerrGeodesic} of a generic photon is
\begin{align}
	\tilde{p}=\frac{1}{2M^2\Gamma}\pa{\frac{\ell\delta+4\omega M(1-\delta)}{2R\Gamma}\pd_T\pm_r\frac{R}{\Gamma}\pa{\ell-\omega M\delta}\pd_R+\omega\beta_s\pd_\theta+\frac{\ell\pa{\frac{1}{\delta}-\frac{3}{2}}-\omega M\pa{2-\frac{5}{2}\delta+\frac{1}{4}\delta^2}}{\Gamma}\pd_\Phi},
\end{align}
where we introduced $\delta(\theta)\equiv\Gamma(\theta)\Lambda(\theta)=\sin^2{\theta}$ and $\beta_s$ was defined in Eq.~\eqref{eq:SourceScreen}. There is a unique corresponding choice of subleading-order polarization obeying the conditions in Eq.~\eqref{eq:SubleadingSymmetryConditions},
\begin{align}
	\tilde{f}=\frac{R}{4M\Gamma(2M\omega-\ell)}\Bigg(&\frac{2\ell\Gamma\cot{\theta}+\omega\pa{M\sin{2\theta}\mp_r\beta_s}}{R}\pd_T+\omega R\pa{\beta_s\pm_rM\sin{2\theta}}\pd_R\\
	&\qquad\pm_r\frac{\ell\pa{\frac{2}{\delta}-3}+\omega M\pa{2-\delta+\delta^2}}{\Gamma}\pd_\theta+\br{\frac{\ell\pa{\frac{2}{\delta}-1}-\omega M(2+\delta)}{\tan{\theta}}\pm_r\omega\beta_s}\pd_\Phi\Bigg),\nonumber
\end{align}
which solves Eqs.~\eqref{eq:Beam} to subleading order in $\epsilon$. Here, note that we cannot impose special-conformal invariance---in fact, $\L_{H_-}\tilde{f}\not\propto\tilde{f}$---but we do continue to impose the symmetries of Kerr, $\L_{H_+}\tilde{f}=\L_{W_0}\tilde{f}=0$. With this choice, we reproduce the expression \eqref{eq:LeadingKappa} for the subleading piece of the Penrose-Walker constant.

\section{Analytic ray-tracing between the near-horizon and asymptotically flat regions}
\label{app:GeodesicEquation}

The polarization vector \eqref{eq:ObservedPolarization} depends explicitly on the polar angle $\theta_s$ at the point of emission. Therefore, in order to determine the polarization profile measured by a distant observer located at $(t_o,r_o,\theta_o,\phi_o)$, we need to find the geodesic connecting a point $(\alpha_o,\beta_o)$ on the observer's screen to the point of emission $(t_s,r_s,\theta_s,\phi_s)$ in the near-horizon region. Without loss of generality, we can set $t_o=\phi_o=0$ and assume that the observer lies in the northern hemisphere, $\theta_o\in\br{0,\pi/2}$, because the entire configuration is stationary, axisymmetric, and reflection-symmetric across the equatorial plane, respectively.

The null geodesic equation in Kerr is reviewed at length in Ref.~\cite{Gralla2018}. The $(r,\theta)$ part of the equation is
\begin{align}
	\label{eq:GeodesicEquation}
	\fint_{r_s}^{r_o}\frac{\ed r}{\pm\sqrt{\mathcal{R}(r)}}=\fint_{\theta_s}^{\theta_o}\frac{\ed\theta}{\pm\sqrt{\Theta(\theta)}},
\end{align}
where the integrals are to be evaluated along the geodesic, with the signs of the integrands chosen to match those of $\ed r$ and $\ed\theta$, respectively. Meanwhile, the $t$ and $\phi$ components of the geodesic equation determine the coordinate time elapsed $\Delta t=t_o-t_s$ and the number of windings about the axis of symmetry $n=\left\lfloor{(\phi_o-\phi_s)}/(2\pi)\right\rfloor$. These quantities are both manifestly real whenever the integrals in Eq.~\eqref{eq:GeodesicEquation} are real, but we do not need to compute them in this problem thanks to the assumed stationarity and axisymmetry of the polarization profile at the source.

The ray-tracing problem has therefore reduced to a two-dimensional calculation in the poloidal plane: given a photon originating from the point $(r_s,\theta_s)$ in the near-horizon region, we must solve Eq.~\eqref{eq:GeodesicEquation} in order to determine the corresponding point $(\alpha_o(\theta_s),\beta_o(\theta_s))$ that the photon hits on the observer's screen, after which we can plug into Eq.~\eqref{eq:ObservedPolarization} to determine the observed polarization at that point.

We are interested in light emitted from the near-horizon region of extreme black holes with $a=M$. In that case, the entire NHEK region of emission is squeezed into the Boyer-Lindquist horizon radius $r=M$. Generic $H=1$ geodesics traverse this region at constant $\theta$, since their leading-order four-momentum is of the form \eqref{eq:H1}, $i.e.$, $p^\theta=0$. Hence, these geodesics emerge from the near-horizon region at $(r_s,\theta_s)$, with $r_s=M$ and $\theta_s$ the angle of emission.

Having fixed the parameters of the problem, it is helpful to introduce new energy-rescaled dimensionless variables\footnote{Note that the variables $(\lambda,q)$ defined here differ from those introduced in Ref.~\cite{Gralla2018}, which were adapted to the study of $H=0$ geodesics. The variable $Q$ is the Carter integral mentioned in footnote \ref{fn:CarterIntegral}.}
\begin{align}
	\label{eq:DimensionlessConstants}
	R=\frac{r-M}{M},\qquad
	\lambda=\frac{\ell}{\omega M},\qquad
	q=\frac{1}{\omega M}\sqrt{k-\pa{\ell-\omega M}^2}=\frac{\sqrt{Q}}{\omega M},\qquad
	\Delta_r=\frac{1}{2}\sqrt{q^2+\pa{\lambda-1}^2}.
\end{align}
Geodesics that pass through the equatorial plane always satisfy $q^2\ge0$, whereas those that do not have $q^2<0$. Nonetheless, the necessary positivity of $\pa{\omega M}^{-2}\Theta(\theta)=q^2+\pa{\lambda-1}^2-\pa{\lambda\csc{\theta}-\sin{\theta}}^2$ guarantees that $\Delta_r$ is real, and moreover that an asymptotic observer situated at angle $\theta_o$ only receives light with
\begin{align}
	\Delta_r\geq\frac{1}{2}\ab{\sin{\theta_o}-\lambda\csc{\theta_o}}.
\end{align}

In terms of these dimensionless variables, the geodesic equation \eqref{eq:GeodesicEquation} now takes the form $I_r=G_\theta$, where
\begin{align}
	I_r=\fint_{R_s}^{R_o}\frac{\ed R}{\pm\sqrt{R^4+4R^3+\pa{7-q^2-\lambda^2}R^2+4(2-\lambda) R+(2-\lambda)^2}},\qquad
	G_\theta=\fint_{\theta_s}^{\theta_o}\frac{\ed\theta}{\pm\sqrt{q^2+\cos^2{\theta}-\lambda^2\cot^2{\theta}}},
\end{align}
with $R_s=0$. The properties of the radial integral as a function of $(\lambda,q)$ determine the points on the observer's screen that can receive light from the near-horizon region of the black hole. In particular, the quartic polynomial 
\begin{align}
	\mathcal{P}(R)=R^4+4R^3+\pa{7-q^2-\lambda^2}R^2+4(2-\lambda)R+(2-\lambda)^2
\end{align}
appearing in the denominator of the integrand of $I_r$ must remain non-negative at all points along the geodesic. 

Consider a photon emerging from the near-horizon region $R_s=0$. In order for it to reach a distant observer at large radius $R_o\to\infty$, it cannot encounter any radial turning points along its trajectory outside of NHEK.\footnote{Photons emitted from a given NHEK radius can still escape to infinity if they are initially ingoing and reach a turning point at a smaller NHEK radius before turning back out. This happens when one of the roots is near zero, or equivalently, only for photons slightly above the superradiant bound $\lambda=2$. Strictly speaking, these photons will appear slightly outside of the shadow, but parametrically close to the NHEKline, which is slightly broadened as a result. These $H=0$ reflected trajectories are analyzed in Ref.~\cite{Gralla2018} but neglected here, as they can only affect the polarization profile along the NHEKline.} Such a turning point occurs at the smallest positive radius $R_t$ where $\mathcal{P}(R)$ becomes negative (i.e. where $\mathcal{P}(R_t)=0$ and $\mathcal{P}'(R_t)<0$). The roots of $\mathcal{P}(R)$ are given by
\begin{subequations}
\begin{align}
	R_1&=-\pa{\Delta_r+1}+\sqrt{\pa{\Delta_r+1}^2+\lambda-2},&
	R_2&=\pa{\Delta_r-1}-\sqrt{\pa{\Delta_r-1}^2+\lambda-2},\\
	R_3&=\pa{\Delta_r-1}+\sqrt{\pa{\Delta_r-1}^2+\lambda-2},&
	R_4&=-\pa{\Delta_r+1}-\sqrt{\pa{\Delta_r+1}^2+\lambda-2}.
\end{align}
\end{subequations}

Since $\mathcal{P}(0)=(2-\lambda)^2\geq0$ and $\mathcal{P}(\pm \infty)=+\infty$, there are two qualitatively different classes of geodesics that can connect to the observer at infinity. The first class consists of geodesics with $(\lambda,q)$ such that all of the roots $R_i$ are complex or negative. In this case, there are no zeroes of $\mathcal{P}(R)$ lying on the contour of integration extending from $R=0$ to $R=\infty$, so the radial integral converges to a finite quantity and the geodesic encounters a finite number of angular turning points according to Eq.~\eqref{eq:GeodesicEquation}. The second class consists of geodesics with $(\lambda,q)$ such that $\mathcal{P}(R)$ develops a double root on the positive real axis. ($R_4$, if real, is always negative, so there can be at most one double root on the positive real axis.) In this case, the radial integral diverges logarithmically, and the geodesic encounters a divergent number of angular turning points as it winds around the black hole. In fact, the second class of geodesics is a limiting case of the first, where complex or negative roots approach and pinch the contour of integration as we vary $(\lambda,q)$. 

It is straightforward to determine the restricted range of $(\lambda,q)$ labeling the geodesics that directly connect the near-horizon region to the observer. First, note that if $\lambda>2$ (that is, if the photon is below the superradiant bound $\omega<\Omega_H\ell$), then all of the roots are real, $R_1$ and $R_3$ are positive, and the photon cannot escape to infinity.\footnote{Conversely, a photon below the superradiant bound coming in from infinity will bounce off the black hole.} We can therefore restrict our attention to the case $\lambda\leq2$. The roots $R_1$ and $R_4$ may then be real but they can never be positive, so the only constraints come from  $R_2$ and $R_3$. For $\Delta_r<1$, these roots (if real)  are negative. However, for $\Delta_r-1\geq\sqrt{2-\lambda}\geq0$, they are real and positive, and the photon cannot escape to infinity.

In summary, a near-horizon photon can directly reach a distant observer at polar angle $\theta_o$ if and only if $\lambda\leq2$ (it is above the superradiant bound), and moreover 
\begin{align}
	\label{eq:ShadowRange}
	\frac{1}{2}\ab{\sin{\theta_o}-\lambda\csc{\theta_o}}-1\leq\Delta_r-1\leq\sqrt{2-\lambda}.
\end{align} 
This condition defines a region on the observer's screen which coincides identically with the shadow cast by an extremal black hole, as can be verified by substituting
\begin{align}
	\lambda=-\frac{\alpha_o\sin{\theta_o}}{M},\qquad
	q=\frac{1}{M}\sqrt{\pa{\alpha_o^2-M^2}\cos^2{\theta_o}+\beta_o^2}, 
\end{align}
and comparing to Eqs.~\eqref{eq:ExtremalShadow}. The interior of this region corresponds to geodesics of the first class discussed above, while near the boundary $\Delta_r-1=\sqrt{2-\lambda}$ the polynomial $\mathcal{P}(R)$ develops at double root at $R=\sqrt{2-\lambda}$. (Note that the shadow's vertical edge \eqref{eq:NHEKline} lies on the line $\lambda=2$, corresponding to the case when the double root emerges at the lower endpoint of integration.) Therefore, light rays that impinge upon the observer's screen near the edge of the shadow must have librated around the black hole an infinite number of times before reaching the asymptotic region.

For the parameter range \eqref{eq:ShadowRange}, the radial integral takes the form
\begin{align}
	I_r=\frac{2}{\sqrt{r_{12}r_{34}}}\br{F\pa{\arcsin\sqrt{\frac{r_{12}}{r_{32}}}\left|\frac{r_{14}r_{32}}{r_{12}r_{34}}\right.}-F\pa{\arcsin\sqrt{\frac{R_3}{R_1}\frac{r_{12}}{r_{32}}}\left|\frac{r_{14}r_{32}}{r_{12}r_{34}}\right.}},
\end{align}
where $r_{ij}\equiv R_i-R_j$ and $F(\phi|x)$ denotes the incomplete elliptic integral of the first kind.

We have seen that light emitted from the NHEK region must always appear within the shadow cast by the black hole on the observer's screen. In order to find the explicit map from the point of emission to the corresponding point on the observer's screen, we still have to solve the $(r,\theta)$ geodesic equation \eqref{eq:GeodesicEquation}. As was the case for the radial integral, one can read off qualitative features of the angular motion from the nature of the roots of the polynomial appearing in the integrand of $G_{\theta}$. After a change of variables $u=\cos^2\theta$, that integrand becomes
\begin{align}
	\sign\pa{\theta-\frac{\pi}{2}}\frac{\ed\theta}{\sqrt{\Theta(\theta)}}=\frac{1}{2}\frac{\ed u}{\sqrt{u(u_+-u)(u-u_-)}},\qquad
	u_{\pm}=\Delta_\theta\pm\sqrt{\Delta_\theta^2+q^2},\qquad
	\Delta_\theta=\frac{1}{2}\pa{1-q^2-\lambda^2}.
\end{align}
Geodesic motion in the angular direction is constrained by the requirement that the argument of the square root remain positive. While we always have $u_+\in\br{0,1}$, there are two qualitatively different behaviors distinguished by the sign of $u_-$. For $q^2>0$, $u_-$ is negative and the geodesic oscillates about the equatorial plane in the angular region $\cos^2\theta\in\br{0,u_+}$. On the other hand, when $q^2<0$, then $u_-$ is also positive and the geodesic is constrained to the angular region $\cos^2\theta\in\br{u_-,u_+}\subset\br{0,1}$, defining a cone either entirely above or entirely below the equatorial plane. We will ignore the boundary case $q=0$, as it corresponds to a measure-zero curve on the observer's screen.

The angular integral $G_\theta$ was evaluated for Kerr geodesics with $q^2>0$ (and generic $a$) in Ref.~\cite{Gralla2018} in terms of $F(\phi|x)$, the incomplete elliptic integral of the first kind. We will adopt the convention
\begin{align}
	F(\phi|x)=\int_0^\phi\frac{\ed t}{\sqrt{1-x\sin^2{t}}},\qquad
	\Psi_j=\arcsin\pa{\frac{\cos{\theta_j}}{\sqrt{u_+}}},\qquad
	\Upsilon_j=\arcsin\sqrt{\frac{\cos^2{\theta_j}-u_-}{u_+-u_-}},
\end{align}
where we also introduced symbols $\Psi_j$ and $\Upsilon_j$ for notational convenience. For $\theta_o\in\br{0,\pi/2}$ in extreme Kerr,
\begin{align}
	\label{eq:AngularIntegral+}
	G_\theta^{q^2>0}=\frac{1}{\sqrt{-u_-}}\cu{2mK\pa{\frac{u_+}{u_-}}\pm_o\br{(-1)^mF\pa{\Psi_s\left|\frac{u_+}{u_-}\right.}-F\pa{\Psi_o\left|\frac{u_+}{u_-}\right.}}},
\end{align}
where $m$ is the number of turning points in the polar motion and $K(x)=F(\pi/2|x)$ denotes the complete elliptic integral of the first kind. Likewise, for $q^2<0$, $\theta_o\in\br{0,\pi/2}$ and $a=M$, one can show that 
\begin{align}
	\label{eq:AngularIntegral-}
	G_\theta^{q^2<0}=\frac{1}{\sqrt{u_-}}\pa{m\pm_o\frac{1-(-1)^m}{2}}K\pa{1-\frac{u_+}{u_-}}\pm_o\frac{1}{\sqrt{u_-}}\br{(-1)^mF\pa{\Upsilon_s\left|1-\frac{u_+}{u_-}\right.}-F\pa{\Upsilon_o\left|1-\frac{u_+}{u_-}\right.}}.
\end{align}

We can now solve the geodesic equation \eqref{eq:GeodesicEquation}, $I_r=G_\theta$, for $\theta_s$. After some technical manipulations, this results in
\begin{align}
	\label{eq:CosTheta}
	\cos{\theta_s}=
	\begin{cases}
		\displaystyle\pm_o\sqrt{u_+}(-1)^m\sn\pa{X_m\bigg|\frac{u_+}{u_-}}&\qquad
		q^2>0,\\
		\displaystyle\sqrt{u_-}\dn\pa{Y_m\bigg|1-\frac{u_+}{u_-}}&\qquad
		q^2<0,
	\end{cases}
\end{align}
where we introduced the Jacobi elliptic functions\footnote{The elliptic function $\sn(u|x)$ inverts the incomplete elliptic integral of the first kind: $\sn\pa{F(\arcsin{\phi}|x)|x}=\phi$. The elliptic function $\dn(u|x)$ satisfies $\dn^2(u|x)+x\sn^2(u|x)=1$. While $\sn(-u|x)=-\sn(u|x)$ is odd in its first argument, $\dn(-u|x)=\dn(u|x)$ is even.} $\sn(u|x)$ and $\dn(u|x)$ and defined the quantities
\begin{subequations}
\begin{align}
	X_m&=\sqrt{-u_-}I_r\pm_oF\pa{\Psi_o\bigg|\frac{u_+}{u_-}}-2mK\pa{\frac{u_+}{u_-}},\\
	Y_m&=\sqrt{u_-}I_r\pm_oF\pa{\Upsilon_o\bigg|1-\frac{u_+}{u_-}}-\pa{m\pm_o\frac{1-(-1)^m}{2}}K\pa{1-\frac{u_+}{u_-}}.
\end{align}
\end{subequations}

Although it is not immediately apparent, both expressions for $\cos{\theta_s}$ are in fact independent of the number $m$ of polar turning points along the trajectory. In order to see this, begin by noting that
\begin{subequations}
\begin{align}
	X_{m+1}&=X_m-2K\pa{\frac{u_+}{u_-}},\\
	Y_{m+1}&=Y_m-\br{1\pm_o(-1)^m}K\pa{1-\frac{u_+}{u_-}}.
\end{align}
\end{subequations}
Next, note that because of the periodicity conditions $\sn(u\pm2K(x)|x)=-\sn(u|x)$ and $\dn(u\pm 2K(x)|x)=\dn(u|x)$, the quantities $(-1)^{m}\sn(X_m|u_+/u_-)$ and $\dn(Y_m|1-u_+/u_-)$ are both invariant under integer shifts in $m$. It follows that $X_m$ and $Y_m$ can be respectively replaced by $X_0$ and $Y_0$ in Eq.~\eqref{eq:CosTheta}, from which the $m$-dependence thereby disappears, as claimed.

Moreover, after some further technical manipulations,\footnote{To further simplify the $q^2<0$ expression and relate it to the $q^2>0$ case, one repeatedly invokes the relation
\begin{align*}
	K(x)-\frac{1}{\sqrt{x}}K\pa{\frac{1}{x}}=-iK(1-x),
\end{align*}
valid for $x=u_+/u_-\ge1$, along with the nontrivial identity
\begin{align*}
	\frac{1}{\sqrt{x}}K\left(\frac{1}{x}\right)-F\pa{\arcsin\sqrt{\frac{u_j}{u_+}}\bigg| x}=iF\pa{\arcsin\sqrt{\frac{u_j-u_-}{u_+-u_-}}\bigg|1-x},\qquad
	x=\frac{u_+}{u_-}.
\end{align*}} the two expressions for $\cos{\theta_s}$ in Eq.~\eqref{eq:CosTheta} can be packaged together into the remarkably simple formula (valid for all $q^2\neq0$)
\begin{align}
	\label{eq:SourceAngle}
	\theta_s=\arccos\br{\pm_o\sqrt{u_+}\sn\pa{\frac{\sqrt{-u_-}}{\sign(-u_-)}I_r\pm_oF\pa{\Psi_o\bigg|\frac{u_+}{u_-}}\bigg|\frac{u_+}{u_-}}}.
\end{align}

While the above expression for $\theta_s$ is independent of the integer $m$, the sign $\pm_s$ does depend on its parity,
\begin{align}
	\pm_s=\sign(p_\theta)|_{R_s}
	=(-1)^m\sign(\beta_o)
	=(-1)^m\pm_o,
\end{align}
and must be determined for each geodesic. If $\theta_o=0$, then $\pm_o$ is ill-defined and we take $\pm_s=(-1)^{m+1}$ by convention.\footnote{When $\theta_o=0$, the $\beta_o$-axis becomes ill-defined because the projection of the axis of symmetry onto the plane perpendicular to the observer's line of sight degenerates to a point. Moreover, the value of $m$ is also ill-defined because the observer sits on a turning point of the incoming photons' trajectories. In drawing the `face-on' case in Fig.~\ref{fig:FaceOnEdgeOn}, we adopted the convention that the direct light should be labeled $m=0$, in which case $\pm_s=(-1)^{m+1}$.}

In practice, to produce polarization plots, we only need to find the boundaries between regions of different $m$. Since the source dependent terms in Eqs.~\eqref{eq:AngularIntegral+}-\eqref{eq:AngularIntegral-} are bounded, as the radial integral grows, more turning points are necessary to satisfy the geodesic equation. Schematically, since the radial integral assumes small values near the center of the shadow and diverges near the boundary, one expects nested regions of increasing $m$ as one moves out towards the edge of the shadow. In the $q^2<0$ case, we find that $m=0$ whenever
\begin{align}
	0<\pm_o\sqrt{u_-}I_r+F\pa{\Upsilon_o\bigg|1-\frac{u_+}{u_-}}<K\pa{1-\frac{u_+}{u_-}},
\end{align}
and $m=1$ otherwise. In the $q^2>0$ case, geodesics with $m$ angular turning points obey
\begin{align}
	(2m-1)K\pa{\frac{u_+}{u_-}}<\sqrt{-u_-}I_r\pm_oF\pa{\Psi_o\bigg|\frac{u_+}{u_-}}<(2m+1)K\pa{\frac{u_+}{u_-}}.
\end{align}
The contours of the regions satisfying this inequality for different $m$ are the boundaries of the regions depicted with different colors in Fig.~\ref{fig:FaceOnEdgeOn}. Mathematically, the existence of these regions can ultimately be traced back to the periodicity of the elliptic function $\sn(u|x)$, which implies that its inverse $F\pa{\arcsin{\sn(u|x)}\big|x}$ is multivalued (not equal to $u$ everywhere). As such, when the formula \eqref{eq:SourceAngle} for $\theta_s$ is substituted back into the geodesic integrals \eqref{eq:AngularIntegral+}-\eqref{eq:AngularIntegral-}, one must choose the appropriate branch of the elliptic integral to satisfy Eq.~\eqref{eq:GeodesicEquation}. This inverse has infinitely many branches labeled by $m$, which correspond to the infinitely many coverings of the horizon in our polarimetric images.

\bibliography{Polarization}
\bibliographystyle{utphys}

\end{document}